\documentclass[manyauthors,nocleardouble,COMPASS]{cernphprep}
\usepackage{amssymb}
\usepackage{amsmath}
\usepackage{amsbsy}
\usepackage{times}
\usepackage{epsfig}
\usepackage{colordvi}
\usepackage{graphicx}
\usepackage{wrapfig,rotating}
\usepackage[numbers, square, comma, sort&compress]{natbib}
\usepackage{hyperref} 
\usepackage{multicol}
\usepackage{booktabs}
\usepackage{multirow}
\usepackage{lineno}
\RequirePackage[T1]{fontenc}

\newcommand{\be}{\begin{equation}}
\newcommand{\ee}{\end{equation}}
\newcommand{\bea}{\begin{eqnarray}}
\newcommand{\eea}{\end{eqnarray}}

\newcommand{\open}{{<\kern -0.3 em{\scriptscriptstyle )}}}
\def\>{\rangle}
\def\<{\langle}
\def\bce{\begin{center}}
\def\ece{\end{center}}

\newcommand{\nslash}{\kern 0.2 em n\kern -0.50em /}
\newcommand{\qslash}{\kern 0.2 em q\kern -0.50em /}
\newcommand{\kslash}{\kern 0.2 em k\kern -0.45em /}
\newcommand{\pslash}{\kern 0.2 em p\kern -0.50em /}
\newcommand{\Sslash}{\kern 0.2 em S\kern -0.50em /}
\newcommand{\Pslash}{\kern 0.2 em P\kern -0.50em /}
\newcommand{\Dslash}{\kern 0.2 em D\kern -0.65em /\kern 0.15em}

\newcommand{\xbj}{x}                   

\makeatletter
\newsavebox\myboxA
\newsavebox\myboxB
\newlength\mylenA

\newcommand*\xoverline[2][0.6]{%
    \sbox{\myboxA}{$\m@th#2$}%
    \setbox\myboxB\null
    \ht\myboxB=\ht\myboxA%
    \dp\myboxB=\dp\myboxA%
    \wd\myboxB=#1\wd\myboxA
    \sbox\myboxB{$\m@th\overline{\copy\myboxB}$}
    \setlength\mylenA{\the\wd\myboxA}
    \addtolength\mylenA{-\the\wd\myboxB}%
    \ifdim\wd\myboxB<\wd\myboxA%
       \rlap{\hskip 0.7\mylenA\usebox\myboxB}{\usebox\myboxA}%
    \else
        \hskip -0.7\mylenA\rlap{\usebox\myboxA}{\hskip 0.5\mylenA\usebox\myboxB}%
    \fi}
\makeatother
\DeclareSymbolFont{letters}     {OML}{cmm}{m}{it}
\DeclareSymbolFont{symbols}     {OMS}{cmsy}{m}{n}
\DeclareSymbolFont{largesymbols}{OMX}{cmex}{m}{n}
\begin{document}
\begin{titlepage}
\PHnumber{2014--013}
\PHdate{\today}

\title{A high-statistics measurement of transverse spin 
effects in dihadron production from 
muon-proton semi-inclusive deep-inelastic scattering}

\Collaboration{The COMPASS Collaboration}
\ShortAuthor{The COMPASS Collaboration}

\begin{abstract}
A measurement of the azimuthal asymmetry in dihadron production in
deep-inelastic scattering of muons on transversely polarised proton (NH$_{3}$)
targets are presented. They provide independent access to the transversity
distribution functions through the measurement of the Collins asymmetry in
single hadron production.  The data were taken in the year $2010$ with the
COMPASS spectrometer using a $160\,\mbox{GeV}/c$ muon beam of the CERN SPS,
increasing by a factor of about three the available statistics of the previously
published data taken in the year $2007$.  The measured sizeable asymmetry is in
good agreement with the published data. An approximate equality of the Collins
asymmetry and the dihadron asymmetry is observed, suggesting a common physical
mechanism in the underlying fragmentation.
\end{abstract}
\vfill
\Submitted{(to be submitted to Phys.~Lett.~B)}
\end{titlepage}

{\pagestyle{empty}
%
%

\section*{The COMPASS Collaboration}
\label{app:collab}
\renewcommand\labelenumi{\textsuperscript{\theenumi}~}
\renewcommand\theenumi{\arabic{enumi}}
\begin{flushleft}
C.~Adolph\Irefn{erlangen},
R.~Akhunzyanov\Irefn{dubna},
M.G.~Alekseev\Irefn{triest_i},
Yu.~Alexandrov\Irefn{moscowlpi}\Deceased,
G.D.~Alexeev\Irefn{dubna}, 
A.~Amoroso\Irefnn{turin_u}{turin_i},
V.~Andrieux\Irefn{saclay},
V.~Anosov\Irefn{dubna}, 
A.~Austregesilo\Irefnn{cern}{munichtu},
B.~Bade{\l}ek\Irefn{warsawu}
% ...
\end{flushleft}

%
%

\begin{Authlist}
\item \Idef{bielefeld}{Universit\"at Bielefeld, Fakult\"at f\"ur Physik, 33501 Bielefeld, Germany\Arefs{f}}
\item \Idef{bochum}{Universit\"at Bochum, Institut f\"ur Experimentalphysik, 44780 Bochum, Germany\Arefs{f}\Arefs{ll}}
\item \Idef{bonniskp}{Universit\"at Bonn, Helmholtz-Institut f\"ur  Strahlen- und Kernphysik, 53115 Bonn, Germany\Arefs{f}}
\item \Idef{bonnpi}{Universit\"at Bonn, Physikalisches Institut, 53115 Bonn, Germany\Arefs{f}}
\item \Idef{brno}{Institute of Scientific Instruments, AS CR, 61264 Brno, Czech Republic\Arefs{g}}
\item \Idef{calcutta}{Matrivani Institute of Experimental Research \& Education, Calcutta-700 030, India\Arefs{h}}
\item \Idef{dubna}{Joint Institute for Nuclear Research, 141980 Dubna, Moscow region, Russia\Arefs{i}}
\item \Idef{erlangen}{Universit\"at Erlangen--N\"urnberg, Physikalisches Institut, 91054 Erlangen, Germany\Arefs{f}}
\item \Idef{freiburg}{Universit\"at Freiburg, Physikalisches Institut, 79104 Freiburg, Germany\Arefs{f}\Arefs{ll}}
\item \Idef{cern}{CERN, 1211 Geneva 23, Switzerland}
\item \Idef{liberec}{Technical University in Liberec, 46117 Liberec, Czech Republic\Arefs{g}}
\item \Idef{lisbon}{LIP, 1000-149 Lisbon, Portugal\Arefs{j}}
\item \Idef{mainz}{Universit\"at Mainz, Institut f\"ur Kernphysik, 55099 Mainz, Germany\Arefs{f}}
\item \Idef{miyazaki}{University of Miyazaki, Miyazaki 889-2192, Japan\Arefs{k}}
\item \Idef{moscowlpi}{Lebedev Physical Institute, 119991 Moscow, Russia}
\item \Idef{munichlmu}{Ludwig-Maximilians-Universit\"at M\"unchen, Department f\"ur Physik, 80799 Munich, Germany\Arefs{f}\Arefs{l}}
\item \Idef{munichtu}{Technische Universit\"at M\"unchen, Physik Department, 85748 Garching, Germany\Arefs{f}\Arefs{l}}
\item \Idef{nagoya}{Nagoya University, 464 Nagoya, Japan\Arefs{k}}
\item \Idef{praguecu}{Charles University in Prague, Faculty of Mathematics and Physics, 18000 Prague, Czech Republic\Arefs{g}}
\item \Idef{praguectu}{Czech Technical University in Prague, 16636 Prague, Czech Republic\Arefs{g}}
\item \Idef{protvino}{State Research Center of the Russian Federation, Institute for High Energy Physics, 142281 Protvino, Russia}
\item \Idef{saclay}{CEA IRFU/SPhN Saclay, 91191 Gif-sur-Yvette, France\Arefs{ll}}
\item \Idef{telaviv}{Tel Aviv University, School of Physics and Astronomy, 69978 Tel Aviv, Israel\Arefs{m}}
\item \Idef{triest_i}{Trieste Section of INFN, 34127 Trieste, Italy}
\item \Idef{triest_u}{University of Trieste, Department of Physics, 34127 Trieste, Italy}
\item \Idef{triest_ictp}{Abdus Salam ICTP, 34151 Trieste, Italy}
\item \Idef{turin_u}{University of Turin, Department of Physics, 10125 Turin, Italy}
\item \Idef{turin_i}{Torino Section of INFN, 10125 Turin, Italy}
\item \Idef{turin_p}{University of Eastern Piedmont, 15100 Alessandria, Italy}
\item \Idef{warsaw}{National Centre for Nuclear Research, 00-681 Warsaw, Poland\Arefs{n} }
\item \Idef{warsawu}{University of Warsaw, Faculty of Physics, 00-681 Warsaw, Poland\Arefs{n} }
\item \Idef{warsawtu}{Warsaw University of Technology, Institute of Radioelectronics, 00-665 Warsaw, Poland\Arefs{n} }
\item \Idef{yamagata}{Yamagata University, Yamagata, 992-8510 Japan\Arefs{k} }
\end{Authlist}
%
%
\vspace*{-\baselineskip}\renewcommand\theenumi{\alph{enumi}}
\begin{Authlist}
\item \Adef{a}{Also at Instituto Superior T\'ecnico, Universidade de Lisboa, Lisbon, Portugal}
\item \Adef{aa}{Also at Department of Physics, Pusan National University, Busan 609-735, Republic of Korea and at Physics Department, Brookhaven National Laboratory, Upton, NY 11973, U.S.A. }
\item \Adef{bb}{Supported by the DFG Research Training Group Programme 1102  ``Physics at Hadron Accelerators''}
\item \Adef{b}{Also at Chubu University, Kasugai, Aichi, 487-8501 Japan\Arefs{k}}
\item \Adef{c}{Also at KEK, 1-1 Oho, Tsukuba, Ibaraki, 305-0801 Japan}
\item \Adef{cc}{Also at Moscow Institute of Physics and Technology, Moscow Region, 141700, Russia}
\item \Adef{y}{present address: National Science Foundation, 4201 Wilson Boulevard, Arlington, VA 22230, United States}
\item \Adef{x}{present address: RWTH Aachen University, III. Physikalisches Institut, 52056 Aachen, Germany}
\item \Adef{e}{Also at GSI mbH, Planckstr.\ 1, D-64291 Darmstadt, Germany}
\item \Adef{f}{Supported by the German Bundesministerium f\"ur Bildung und Forschung}
\item \Adef{g}{Supported by Czech Republic MEYS Grants ME492 and LA242}
\item \Adef{h}{Supported by SAIL (CSR), Govt.\ of India}
\item \Adef{i}{Supported by CERN-RFBR Grants 08-02-91009 and 12-02-91500}
\item \Adef{j}{\raggedright Supported by the Portuguese FCT - Funda\c{c}\~{a}o para a Ci\^{e}ncia e Tecnologia, COMPETE and QREN, Grants CERN/FP/109323/2009, CERN/FP/116376/2010 and CERN/FP/123600/2011}
\item \Adef{k}{Supported by the MEXT and the JSPS under the Grants No.18002006, No.20540299 and No.18540281; Daiko Foundation and Yamada Foundation}
\item \Adef{l}{Supported by the DFG cluster of excellence `Origin and Structure of the Universe' (www.universe-cluster.de)}
\item \Adef{ll}{Supported by EU FP7 (HadronPhysics3, Grant Agreement number 283286)}
\item \Adef{m}{Supported by the Israel Science Foundation, founded by the Israel Academy of Sciences and Humanities}
\item \Adef{n}{Supported by the Polish NCN Grant DEC-2011/01/M/ST2/02350}
\item [{\makebox[2mm][l]{\textsuperscript{*}}}] Deceased
\end{Authlist}

\clearpage}

\section {Introduction}

\noindent
The quark structure of the nucleon can be characterised by parton distribution
functions (PDFs) for each quark flavour~\cite{Jaffe:1991kp}.  If the quark
intrinsic transverse momentum $\textbf{\textit{k}}_{T}$ is integrated over,
there remain at twist-two level three PDFs depending on the Bjorken scaling
variable $x$ and the negative square of the four-momentum transfer $Q^{2}$,
which exhaust the information on the partonic structure of the nucleon
~\cite{Jaffe:1991ra,Kotzinian:1994dv,Mulders:1995dh,Barone:2001sp}.  The
spin-independent distribution $f_{1}^{q}$ and the helicity distribution
$g_{1}^{q}$ have been measured with good accuracy.  However, up to ten years
ago nothing was known about the transverse spin distribution $h^{q}_{1}$,
often referred to as transversity, which describes the probability difference of
finding a quark $q$ polarised parallel or antiparallel to the spin of a
transversely polarised nucleon. This distribution is difficult to measure, since
it is related to soft processes correlating quarks with opposite chirality,
making it a chiral-odd function~\cite{Jaffe:1991kp}. As a result, transversity
can only be accessed through observables in which it appears coupled to a second
chiral-odd object in order to conserve chirality. Thus it does not contribute to
inclusive deep-inelastic scattering (DIS) at leading twist.  In semi-inclusive
deep-inelastic scattering (SIDIS) reactions the chiral-odd partners of the
transversity distribution function are fragmentation functions (FFs), which
describe the spin-dependent hadronisation of a transversely polarised quark $q$
into hadrons. For a recent review see Ref.~\cite{Barone:2010zz}. Up to now,
most of the information on transversity came from the Collins asymmetry measured
in single hadron asymmetries
~\cite{PhysRevLett.94.012002,PhysRevLett.94.202002,Alekseev:2010rw,Airapetian201011}
and used in global analyses (e.g.~\cite{Anselmino:2008jk}).  

A complementary approach is to measure dihadron production in leptoproduction in
SIDIS on transversely polarised nucleon, $l\,N^{\uparrow} \, \rightarrow \,
l^{\prime}\,h^{+}\,h^{-}\,X$ with both hadrons produced in the current
fragmentation region
~\cite{Collins:1993kq,Artru:1995zu,Jaffe:1997hf,Radici:2001na}.  In this
reaction appears a new chiral-odd fragmentation function, the dihadron
Fragmentation Function (DiFF) $H_{1}^{\sphericalangle}$, which describes the
spin-dependent part of the fragmentation of a transversely polarised quark into
a pair of unpolarised hadrons describing a correlation of quark transverse spin
with normal pseudo-vector to the dihadron momenta plane (the
handedness)~\cite{Efremov1992394}.  The transverse polarisation of the
fragmenting quark is correlated with the relative momentum of the two hadrons,
which gives rise to a transverse, target-spin-dependent azimuthal asymmetry
around the virtual-photon direction, with respect to the lepton scattering
plane.  In this case, the sum of the total transverse momenta of the final state
hadrons can be integrated over, leaving only the relative momentum of the two
hadrons.  This avoids the complexity of transverse-momentum-dependent
convolution integrals as in the analysis of single hadron production utilising
the Collins effect and the analysis can be performed using collinear
factorisation~\cite{Boer:2008mz,Bacchetta:2008wb}. Here, the evolution equations
are known at next-to-leading order ~\cite{Ceccopieri:2007ip}, so that results
from $e^{+}e^{-}$ scattering and SIDIS can be correctly connected, making it a
theoretically clean way to extract transversity using presently existing
facilities \cite{Boer:2008mz}. The properties of the DiFFs are described in
detail in Refs.~\cite{Collins:1993kq,Artru:1995zu,Jaffe:1997hf,Radici:2001na,Bianconi:1999cd,
  Bacchetta:2002ux,Bacchetta:2003vn,Bacchetta:2006un}. 

First evidence for an azimuthal asymmetry in leptoproduction of $\pi^{+}\pi^{-}$
pairs was published by HERMES, using a transversely polarised hydrogen
target~\cite{Airapetian:2008sk}. The DiFFs were first measured in $e^{+}e^{-}$
reactions by Belle~\cite{Vossen:2011fk} and BaBar\cite{TheBABAR:2013yha}. These
measurements indicate a sizeable $u$ quark transversity distribution -- as
already known from the measurements of the Collins asymmetry
~\cite{Alekseev:2010rw,Adolph:2012sn,PhysRevLett.94.012002} -- and non-vanishing
DiFFs~\cite{Bacchetta:2011ip,PhysRevLett.94.012002}.\newline Recently, COMPASS
published results on dihadron asymmetry obtained from the data collected using
transversely polarised deuteron ($^{6}$LiD) and proton (NH$_{3}$) targets in the
years $2002$-$2004$ and $2007$, respectively~\cite{Adolph:2012nw}.  Due to the
large acceptance of the COMPASS spectrometer and the large muon momentum of
$160\,\mbox{GeV}/c$, results with high statistics were obtained covering a large
kinematic range in $x$ and $M_{h^{+}h^{-}}$, the invariant mass of the
dihadron. Sizeable asymmetries were measured on the proton target while on the
deuteron target only small asymmetries were observed.  These results indicate
non-vanishing $u$ quark transversity and DiFFs, as well as a cancellation of the
contributions of $u$ and $d$ quark transversities in the deuteron.  Using these
data sets in conjunction with the Belle data, a first parametrisation of the $u$
and $d$ quark transversities was performed based on a collinear
framework~\cite{Bacchetta:2012ty}.  The same procedure was applied to directly
extract $u$ and $d$ quark transversities in the same $x$ bins as used to obtain
the COMPASS proton and deuteron results~\cite{Elia:2012}.  In this Letter, the
dihadron azimuthal asymmetries measured from the data collected in $2010$ with a
transversely polarised proton target (NH$_{3}$, as in 2007) are presented.  The
statistics accumulated in this data taking period increases the total available
statistics on proton by a factor of four.

\section {Theoretical Framework}
\noindent
Here, only a short summary of the theoretical framework is given.  For a more
detailed view, we recommend the references given above and our recent
paper~\cite{Adolph:2012nw} on the same topic.\newline At leading twist and after
integration over total transverse momenta, the cross section of semi-inclusive
dihadron leptoproduction on a transversely polarised target is given as a sum of
a spin-independent and a spin-dependent part
\cite{Bacchetta:2002ux,Bacchetta:2003vn}:
\bea
\frac{d^7\,\sigma_{UU}}{d\cos\theta\,dM_{h^{+}h^{-}}^2\,d\phi_R\,dz\,dx\,dy\,d\phi_S} &= & \frac{\alpha^2}{2\pi Q^2 y}\,\left(1-y+\frac{y^2}{2}\right)   \\
        & &   \times  \sum_q e_{q}^{2} f_1^q(\xbj)\, D_{1,q}\bigl(z, M_{h^{+}h^{-}}^2, \cos \theta\bigr) , \nonumber \\
\frac{d^7\,\sigma_{UT}}{d\cos\theta\,dM_{h^{+}h^{-}}^2\,d\phi_R\,dz\,dx\,dy\,d\phi_S} &= &\frac{\alpha^2}{2\pi Q^2 y}\, S_{\perp}^{} \,(1-y)  \\ 
 & &  \times \sum_q e_{q}^{2}  \frac{|\textbf{\textit{p}}_{1} - \textbf{\textit{p}}_{2}| }{2 M_{h^{+}h^{-}}} \, \sin \theta \, \sin\phi_{RS}^{}\,
   h_1^q(\xbj)\,H_{1,q}^{\sphericalangle}\bigl(z, M_{h^{+}h^{-}}^2, \cos \theta\bigr) .\nonumber
\label{eq:crossOT}
\eea
Here, the sums run over all quark and antiquark flavours $q$,
$\textbf{\textit{p}}_{1}$ and $\textbf{\textit{p}}_{2}$ denote the three-momenta
of the two hadrons of the dihadron, where the subscript $1$ always refers to the
positive hadron in this analysis. The first subscript ($U$) indicates an
unpolarised beam and the second ($U$ or $T$), an unpolarised and transversely
polarised target, respectively. Note that the contribution from a longitudinally
polarised beam and a transversely polarised target, $\sigma_{LT}$, is neglected
in this analysis since it exhibits a different azimuthal angle and is suppressed
by a factor of $1/Q$~\cite{Bacchetta:2003vn}. The fine-structure constant is
denoted by $\alpha$, $y$ is the fraction of the muon energy transferred to the
virtual photon, $D_{1,q}(z, M^{2}_{h^{+}h^{-}}, \cos \theta)$ is the
spin-independent dihadron fragmentation function for a quark of flavour $q$,
$H^{\sphericalangle}_{1,q}(z,M^{2}_{h^{+}h^{-}},\cos \theta)$ is the
spin-dependent DiFF and $z_{1}$, $z_{2}$ are the fractions of the virtual-photon
energy carried by these two hadrons with $z = z_{1} + z_{2}$.  The symbol
$S_\perp$ denotes the component of the target spin vector $\textbf{\textit{S}}$
perpendicular to the virtual-photon direction, and $\theta $ is the polar angle
of one of the hadrons -- commonly the positive one -- in the dihadron rest frame
with respect to the dihadron boost axis. The azimuthal angle $\phi_{RS}$ is
defined as
\begin{figure}
\center
\includegraphics[width=.55\textwidth]{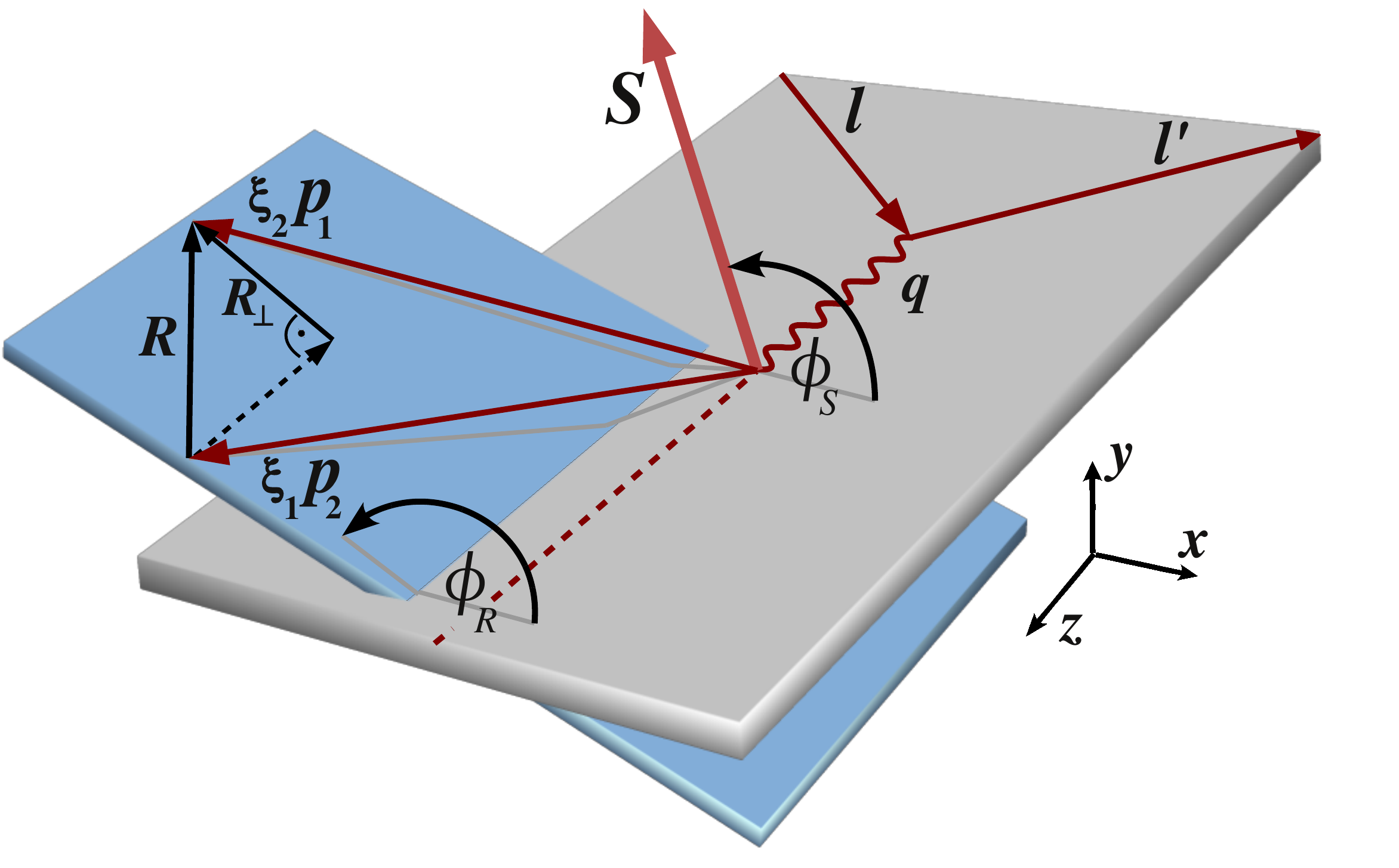}
\caption{
Schematic view of the azimuthal angles $\phi_{R}$ and $\phi_{S}$ for dihadron
production in deep-inelastic scattering, where $\textbf{\textit{l}}$,
$\textbf{\textit{l}}^{\prime}$, $\textbf{\textit{q}}$ and
$\textbf{\textit{p}}_{i}$ are the three-momenta of beam, scattered muon, virtual
photon and hadrons respectively, in the $\gamma^{*}$-nucleon system. Note that
the azimuthal plane is defined by the directions of the relative hadron momentum
and the virtual photon.
\label{angles}
}
\end{figure}
\be
\phi_{RS}  =  \phi_{R}  -  \phi_{S^{\prime}}  =   \phi_{R}  +  \phi_{S}  -  \pi \, ,
\label{def_2h}
\ee
where $\phi_{S}$ is the azimuthal angle of the initial nucleon spin and
$\phi_{S^{\prime}}$ is the azimuthal angle of the spin vector of the fragmenting
quark with $\phi_{S^{\prime}} = \pi - \phi_{S}$ (Fig.~\ref{angles}). The
azimuthal angle $\phi_{R}$ is defined by
\be
\phi_R = \frac{(\textbf{\textit{q}} \times \textbf{\textit{l}}) \cdot \textbf{\textit{R}}}{|(\textbf{\textit{q}} \times \textbf{\textit{l}}) \cdot \textbf{\textit{R}}|} \arccos 
\left( \frac{(\textbf{\textit{q}} \times \textbf{\textit{l}}) \cdot (\textbf{\textit{q}} \times \textbf{\textit{R}})}{|\textbf{\textit{q}} \times \textbf{\textit{l}}||\textbf{\textit{q}} \times \textbf{\textit{R}}|}\right) \, ,
\label{phi_rs}
\ee
where $\textbf{\textit{l}}$ is the incoming lepton momentum,
$\textbf{\textit{q}}$ the virtual-photon momentum and $\textbf{\textit{R}}$ the
relative hadron momentum ~\cite{Artru:1995zu,Artru:2002pua} given by
\be
 \textbf{\textit{R}} =\frac{z_{2} \textbf{\textit{p}}_{1}-z_{1} \textbf{\textit{p}}_{2}}{z_{1} + z_{2}} =\mathrel{\mathop:} \xi_2 \textbf{\textit{p}}_{1}-\xi_1 \textbf{\textit{p}}_{2} \, .
\ee
The number $N_{h^{+}h^{-}}$ of pairs of oppositely charged hadrons produced on a
transversely polarised target can be written as
\be
N_{h^{+}h^{-}}(x,\,y,\,z,\,M^{2}_{h^{+}h^{-}},\,\cos \theta,\,\phi_{RS}) ~ 
\varpropto
 \sigma_{UU} \left( 1 + f(x,y) \, P_T \, D_{nn}(y) \, A_{UT}^{\sin \phi_{RS}} \sin \theta  \sin \phi_{RS} \right) \, ,
\label{h+h-_produced}
\ee
omitting luminosity and detector acceptance. Here, $P_{T}$ is the transverse
polarisation of the target protons and $\textstyle D_{nn}(y) = \frac{1-y}{1 - y
  + y^{2}/2}$ the transverse-spin-transfer coefficient, while $f(x,y)$ is the
target polarisation dilution factor calculated for semi-inclusive reactions
depending on kinematics.  It is given by the abundance-weighted ratio of the
total cross section for scattering on polarisable protons to that for scattering
on all nuclei in the target. The dependence of the dilution factor on the hadron
transverse momenta appears to be weak in the kinematic range of the COMPASS
experiment. Dilution due to radiative events is taken into account by the ratio
of the one-photon exchange cross section to the total cross section.  For
$^{14}$NH$_{3}$, $f$ contains corrections for the polarisation of the spin-$1$
$^{14}$N nucleus.

\noindent The asymmetry 
\be
A_{UT}^{\sin \phi_{RS}} = \frac{|\textbf{\textit{p}}_{1} - \textbf{\textit{p}}_{2}| }{2 M_{h^{+}h^{-}}} \, 
\frac {\sum_q e_{q}^{2} \cdot h_{1}^{q}(x) \cdot H^{\sphericalangle}_{1,q}(z,\,M^{2}_{h^{+}h^{-}},\,\cos \theta)}
 {\sum_q e_{q}^{2} \cdot f_{1}^{q}(x) \cdot D_{1,q}(z,\,M^{2}_{h^{+}h^{-}},\,\cos \theta)}
\label{2h_A_UT}
\ee
is then proportional to the product of the transversity distribution function
and the spin-dependent dihadron fragmentation function, summed over the quark
and antiquark flavours.
\section {Experimental Data and Analysis}
\noindent
The analysis presented in this Letter is performed using data taken in the year
$2010$ with the COMPASS spectrometer~\cite{Abbon:2007pq}, which was obtained by
scattering positive muons of $160\,\mbox{GeV}/c$ produced from the $\mbox{M}2$
beamline of CERNs SPS off a transversely polarised solid-state NH$_{3}$
target. Details on data taking, data quality, event selection and analysis can
be found in Refs.~\cite{Adolph:2012sn,Adolph:2012nw}.
\begin{figure}[t!]
\center
\includegraphics[width=0.55\textwidth]{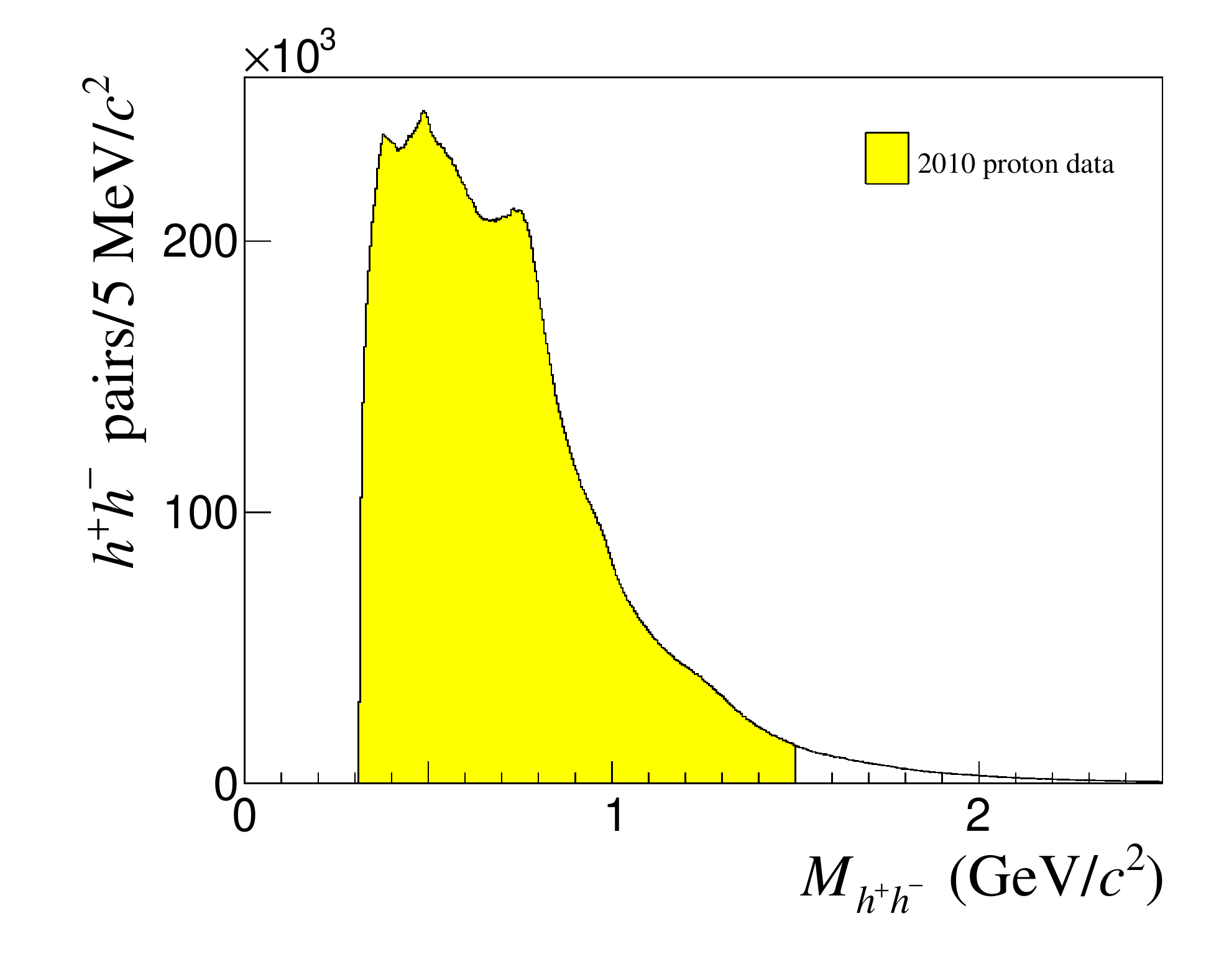}
\caption{
Invariant mass distributions of the final samples. The cut $M_{h^{+}h^{-}} <
1.5\,\mbox{GeV}/c^2$ is indicated.  The $K^{0}$, $\rho$ and $f_{1}$ resonances
are visible.
\label{mass}}
\end{figure}

The beam muons are naturally polarised with an average longitudinal polarisation
of about $0.8$ with a relative uncertainty of $5\,\%$. The average dilution
factor for NH$_{3}$ is $\langle f \rangle \sim 0.15$ and the average
polarisation is $\langle P_T \rangle \sim 0.8$.  The same target as in the year
$2007$ was used.  It consisted of three cylindrical cells with different
orientations of the polarisation vector. In order to compensate acceptance
effects the polarisation was destroyed and built up in reversed direction every
four to five days, for a total of $12$ data-taking sub-periods.\newline For the
analysis, events with incoming and outgoing muons and at least two reconstructed
hadrons from the reaction vertex inside the target cells are selected. Equal
flux through the whole target is obtained by requiring that the extrapolated
beam track crosses all three cells. In order to select events in the DIS regime,
cuts are applied on the squared four-momentum transfer, $Q^2 >
1\,(\mbox{GeV}/c)^2$, and on the invariant mass of the final hadronic state, $W
> 5\,\mbox{GeV}/c^2$. Furthermore, the fractional energy transfer to the virtual
photon is required to be $y > 0.1$ and $ y < 0.9$ to remove events with poorly
reconstructed virtual photon energy and events with large radiative corrections,
respectively.

The dihadron sample consists of all combinations of oppositely charged hadrons
originating from the reaction vertex. Hadrons produced in the current
fragmentation region are selected requiring $z > 0.1$ for the fractional energy
and $x_{F} > 0.1$ of each hadron.  Exclusive dihadron production is suppressed
by requiring the missing energy $\textstyle E_{miss} = \bigl( (P + q - p_{1} -
p_{2})^{2} - m_{P}^{2}\bigr)/(2\,m_{P})$ to be greater than $3.0\,\mbox{GeV}$,
where $P$ is the target protons four-momentum and $m_{P}$ its mass. As the
azimuthal angle $\phi_R$ is only defined for non-collinear vectors
$\textbf{\textit{R}}$ and $\textbf{\textit{q}}$, a minimum value is required on
the component of $\textbf{\textit{R}}$ perpendicular to $\textbf{\textit{q}}$,
$|\textbf{\textit{R}}_{\perp}| > 0.07\,\mbox{GeV}/c$.  After all cuts, $3.5
\times 10^7$ $h^{+}h^{-}$ combinations remain.  Figure~\ref{mass} shows the
invariant mass distributions of the dihadron system, always assuming the pion
mass for each hadron.  A cut of $M_{h^{+}h^{-}} < 1.5\,\mbox{GeV}/c^2$ is
applied in order to allow for the analysis of the data suggested
by~\cite{Bacchetta:2002ux}, where both the spin-dependent and spin-independent
dihadron fragmentation functions are expanded in terms of Legendre polynomials
of $\cos \theta$. While removing only a negligible part of the data, this cut
allows for a convenient restriction to relative $s$- and $p$-waves in this
analysis.

In the analysis we extract the product $A = \langle A_{UT}^{\sin \phi_{RS}} \,
\sin \theta \rangle$, integrated over the angle $\theta$.  For a detailed
discussion we refer to Ref.~\cite{Adolph:2012nw}.  It is important to stress
that in the COMPASS acceptance the opening angle $\theta $ peaks close to
$\pi/2$ with $\langle \sin \theta \rangle = 0.94$ and the $\cos \theta$
distribution is symmetric around zero.  In order to allow for a detailed
consideration of the expansion mentioned above, the mean values of all three
relevant distributions ($\sin \theta$, $\cos \theta$ and $\cos^2 \theta$) for
the individual kinematic bins can be found on HEPDATA~\cite{HEPDATA}.  The
asymmetry is evaluated in kinematic bins of $x$, $z$ or $M_{h^{+}h^{-}}$, while
always integrating over the other two variables.  As estimator the extended
unbinned maximum likelihood function in $\phi_R$ and $\phi_S$ is used, already
described in Ref.~\cite{Adolph:2012nw}.

In order to avoid false asymmetries, care was taken to select only such data for
the analysis for which the spectrometer performance was stable in consecutive
periods of data taking.  This was ensured by extensive data quality tests
described in detail in Ref.~\cite{Adolph:2012sn}. The remaining data sample was
carefully scrutinised for a possible systematic bias in the final asymmetry.
Here, the two main sources for uncertainties are false asymmetries, which can be
evaluated by combining data samples with same target spin orientation, and
effects of acceptance, which can be evaluated by comparing sub-samples
corresponding to different ranges in the azimuthal angle of the scattered muon.
No significant systematic bias could be found and the results from all $12$
sub-periods of data taking proved to be compatible.  Therefore, an upper limit
was estimated comparing the results of the systematic studies to expected
statistical fluctuations.  The resulting systematic uncertainty for each data
point amounts to about $75\,\%$ of the statistical uncertainty.  An additional
scale uncertainty of $2.2\,\%$ accounts for uncertainties in the determination
of target polarisation and target dilution factor calculated for semi-inclusive
reactions~\cite{Alekseev:2010hc}.

\section {Results}
\begin{figure}[!t]
\begin{center}
\includegraphics[width=0.85\textwidth]{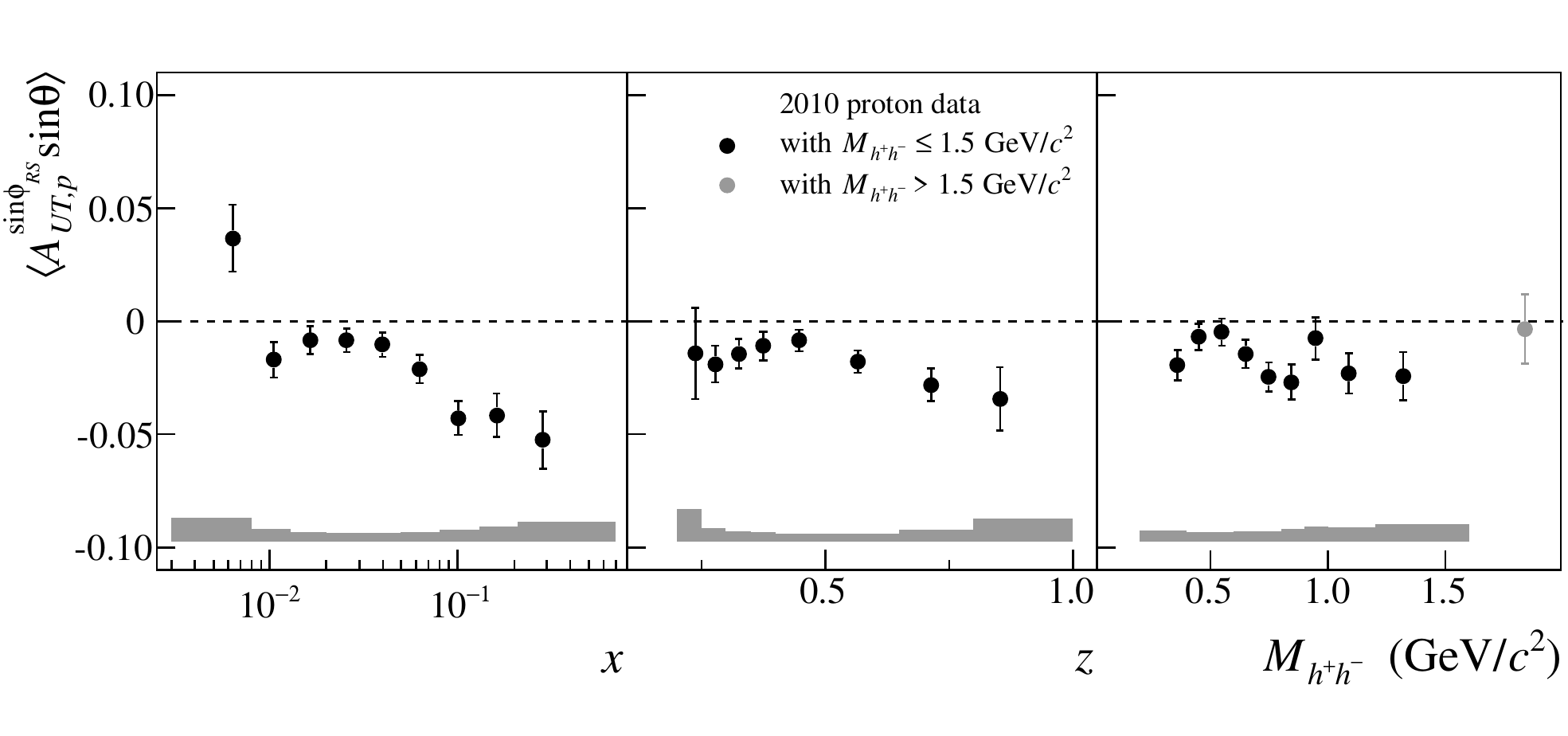}
\caption{Proton asymmetry, integrated over the angle $\theta$, as a function of $x$, $z$ 
and $M_{h^{+}h^{-}}$, for the data taken with the proton (NH$_{3}$) target in the year $2010$.
The grey bands indicate the systematic uncertainties. The last bin in $M_{h^{+}h^{-}}$ 
contains events which were removed from the sample used for results shown as a function
of $x$ and $z$.}
\label{pic:final_asyms_sys_p_2010}
\end{center}
\end{figure}

\begin{figure}[!t]
\begin{center}
\includegraphics[width=0.85\textwidth]{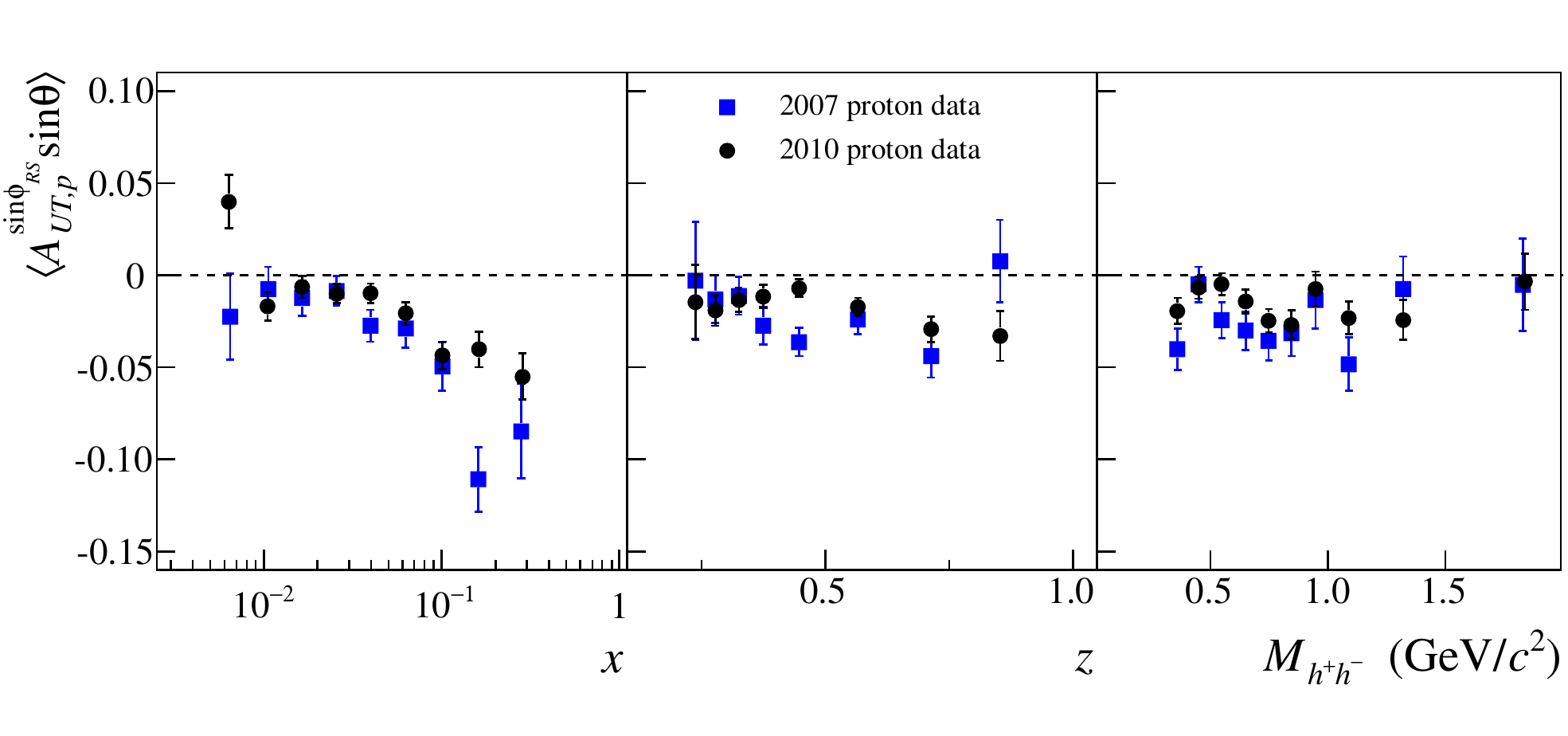}
\caption{Comparison of the asymmetry obtained from the data taken in the years
  $2007$ and $2010$, integrated over the angle $\theta$, as a function of $x$,
  $z$ and $M_{h^{+}h^{-}}$, respectively.}
\label{pic:final_asyms_sys_p_comp}
\end{center}
\end{figure}

\begin{figure}[!t]
\begin{center}
\includegraphics[width=0.85\textwidth]{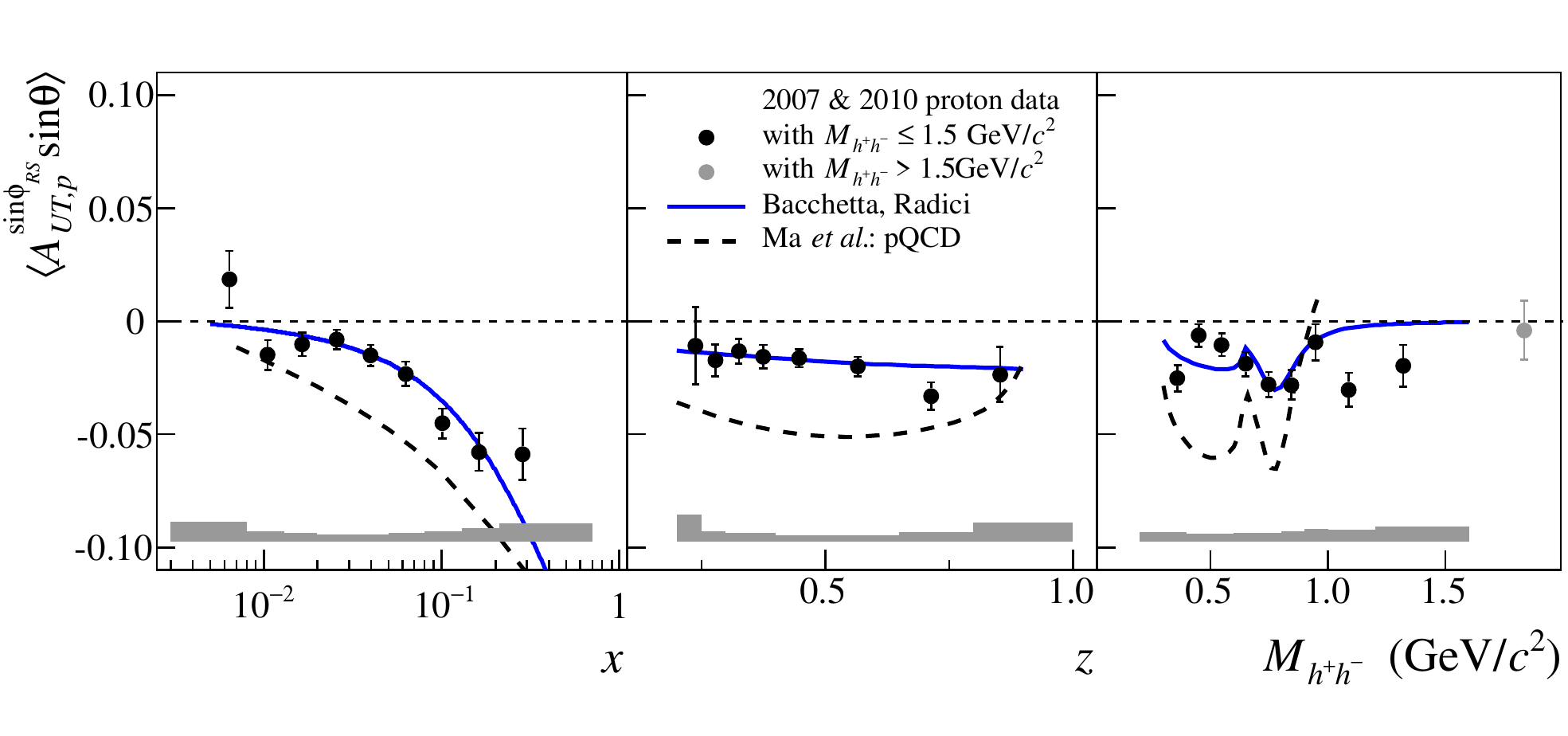}
\includegraphics[width=0.85\textwidth]{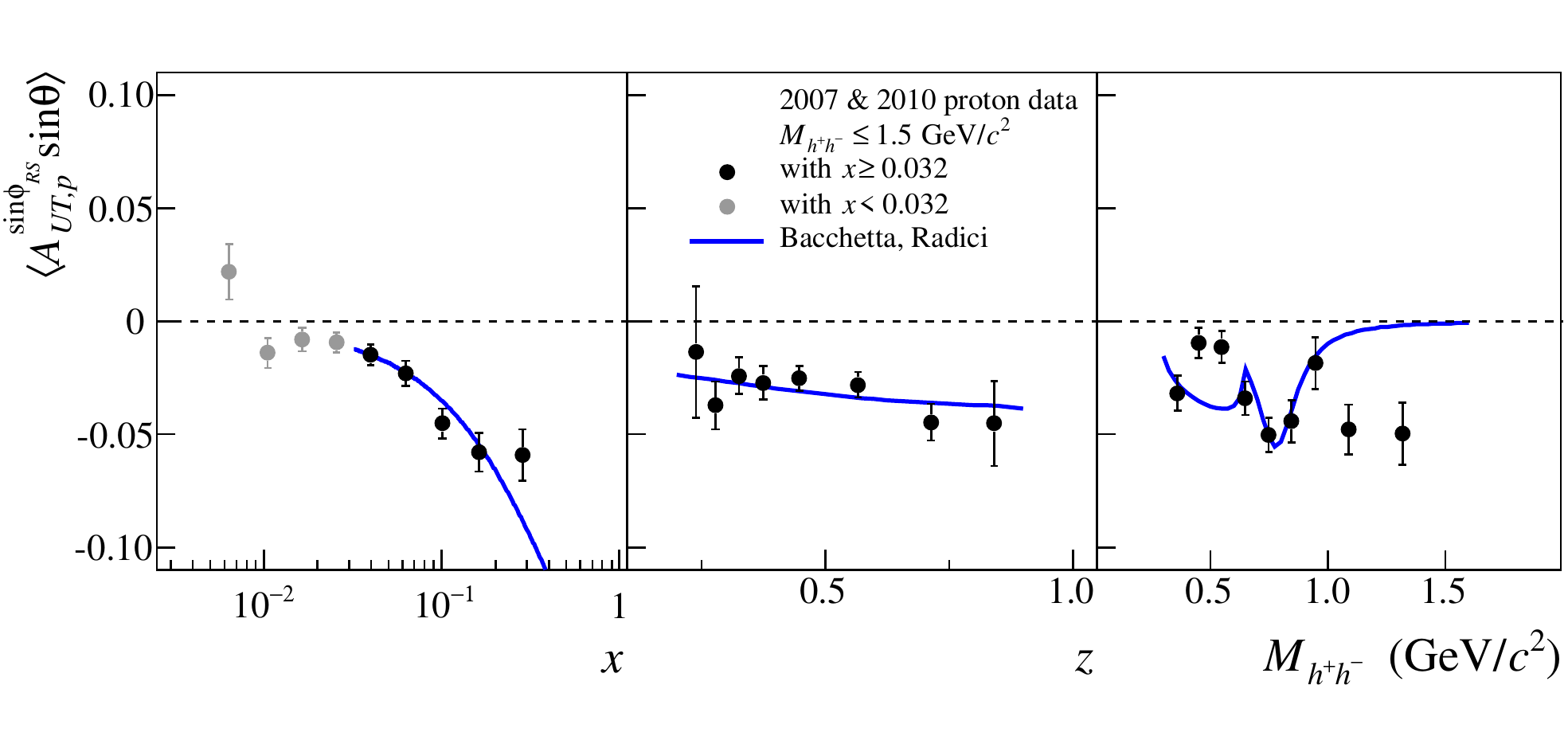}
\caption{Proton asymmetry, integrated over the angle $\theta$, as a function of
  $x$, $z$ and $M_{h^{+}h^{-}}$, for the combined data taken with the proton
  (NH$_{3}$) target in the years $2007$ and $2010$ (top plot).  The grey bands
  indicate the systematic uncertainties.  The bottom plot shows the same data
  for the valence quark region ($x \ge 0.032$). The curves in the upper plots show
  predictions~\cite{Bacchetta:xxx,She:2007ht} made using the transversity
  functions extracted in Ref.~\cite{Anselmino:2008jk} (solid lines) or a pQCD
  based counting rule analysis (dotted lines). The cureves in the lower plots
  show the predictions of~\cite{Bacchetta:xxx} in the same $x \ge 0.032$
  region. Note that the sign of the original predictions was changed to
  accommodate the phase $\pi$ in the definition of the angle $\phi_{RS}$ used in
  the COMPASS analysis.} 
\label{pic:final_asyms_sys_p_2007_2010}
\end{center}
\end{figure}

\noindent
The obtained asymmetry is shown in Fig.~\ref{pic:final_asyms_sys_p_2010} as a
function of $x$, $z$ and $M_{h^{+}h^{-}}$. Large negative asymmetry amplitudes
are observed in the high $x$ region, which implies that both, the transversity
distributions and the spin-dependent dihadron fragmentation functions do not
vanish. Over the measured range of the invariant mass $M_{h^{+}h^{-}}$ and $z$,
the asymmetry is negative and shows no strong dependence on these
variables. Figure~\ref{pic:final_asyms_sys_p_comp} shows the comparison of the
present results to the previously published COMPASS results on the proton target
from $2007$ data~\cite{Adolph:2012nw}.
The results obtained from the data of $2010$ have significantly smaller
statistical uncertainties then the previous results from $2007$ data and both
are in good agreement (CL of 25\%).  Figure~\ref{pic:final_asyms_sys_p_2007_2010}
(top) shows the final result obtained by combining both data sets together with
predictions from model calculations~\cite{Bacchetta:xxx,She:2007ht}.  The
bottom plot shows the same data with a cut on the quark valence region ($x >
0.03$) enhancing the observed signal as a function of $z$ and $M_{h^{+}h^{-}}$.
In comparison to the published HERMES results~\cite{Airapetian:2008sk}, the
results on the proton target presented in this work show higher statistics and
cover a larger kinematic range in $x$ and $M_{h^{+}h^{-}}$.
In the theoretical
approach~\cite{Bacchetta:2002ux,Bacchetta:2003vn,Bacchetta:2006un}, all dihadron
fragmentation functions for di-pion production were calculated in the framework
of a spectator model for the fragmentation process. Predictions were made for
the DiFF $H^{\sphericalangle}_{1}$ as well as for the $s$- and $p$-wave
contributions to the spin-independent fragmentation functions $D_1$ and in
Ref.~\cite{Bacchetta:2006un} the expected asymmetry for COMPASS were calculated
assuming different models for the transversity distributions. Recently, these
parametrisations of the dihadron fragmentation functions from
Ref.~\cite{Bacchetta:2006un} were also used together with the transversity
distributions extracted from single hadron production~\cite{Anselmino:2008jk} to
make predictions for both proton and deuteron targets in the kinematic range
covered by COMPASS.  The calculated asymmetry is shown as solid blue lines in
Fig.~\ref{pic:final_asyms_sys_p_2007_2010} (top and bottom). The latter adapted
for the cut in $x$, shows a good agreement of these predictions with our
data. Significant asymmetry amplitudes are predicted and the $x$ dependent shape
is well described, as well as for the dependence on $z$ in the case of the
calculations by Bacchetta \textit{et al.}. A good agreement in terms of the
$M_{h^{+}h^{-}}$ dependence is only in the mass region of the $\rho$ meson;
%
no optimization of parameters in the calculation of the dihadron fragmentation
function to extend the agreement over a larger $M_{h^{+}h^{-}}$ region (as e.g.,
the fraction of the $\omega$ to 3$\pi$ decay in the $s-p$ interference) was
performed by the authors.  The prediction of Ma \textit{et
  al.}~\cite{She:2007ht} (dashed lines in
Fig.~\ref{pic:final_asyms_sys_p_2007_2010} (top)) uses the parametrisations
of~\cite{Bacchetta:2006un} for the dihadron fragmentation, together with a model
for the transversity distributions, based on a pQCD counting rule analysis. This
prediction describes the main trend of the data but tends to overestimate the
measured asymmetry.
\begin{figure}[!t]
\begin{center}
\includegraphics[width=0.50\textwidth]{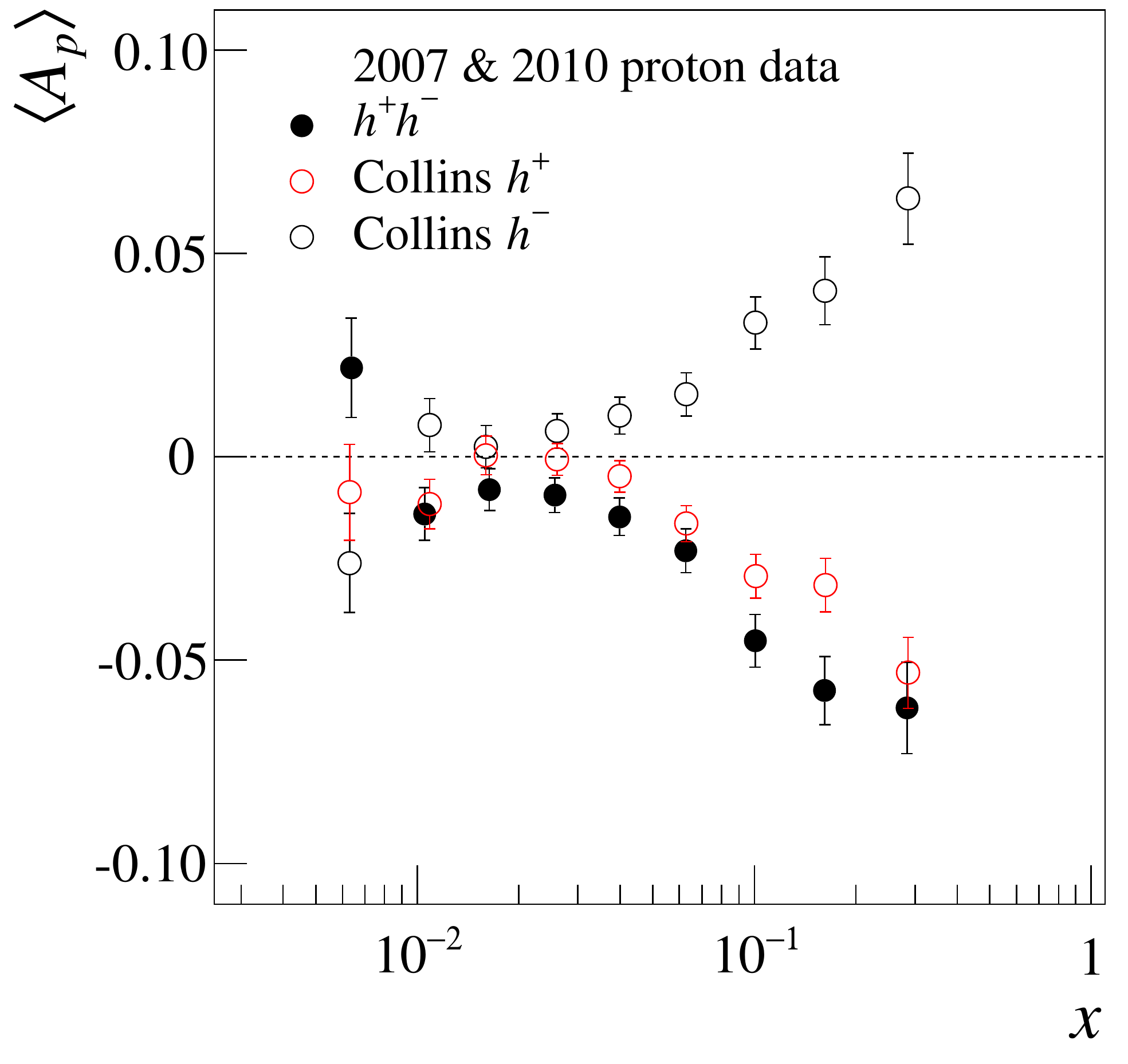}
\caption{Comparison of the asymmetry \textit{vs.} $x$ obtained in the analysis of
  dihadron production to the corresponding Collins asymmetry for the
  combined $2007$ and $2010$ data.
 }
\label{pic:2h_collins}
\end{center}
\end{figure}
\section{Comparing the dihadron asymmetry and the Collins asymmetry}
\noindent
There is a striking similarity among the Collins asymmetry for positive and for
negative hadrons~\cite{Adolph:2012sn} and the dihadron asymmetry as functions of
$x$, as clearly shown in Fig.~\ref{pic:2h_collins}, where the combined results
from the $2007$ and $2010$ COMPASS runs are presented.  First, there is a mirror
symmetry between the Collins asymmetry for positive and for negative hadrons,
the magnitude of the asymmetry being essentially identical and the sign being
opposite.  This symmetry has been phenomenologically described in terms of
opposite signs of $u$ and $d$ quark transversity distributions with almost equal
magnitude and opposite sign for favoured and unfavoured Collins fragmentation
functions~\cite{Anselmino:2008jk}.

The new results show that the values of the
dihadron asymmetry are slightly larger in magnitude, but very close to the
values of the Collins asymmetry for positive hadrons and to the mean of the
values of the Collins asymmetry for positive and negative hadrons, after
changing the sign of the asymmetry of the negative hadrons. The hadron samples
on which these asymmetries are evaluated are
different~\cite{Adolph:2012nw,Adolph:2012sn} since at least one hadron with
$z>0.2$ is required to evaluate the Collins asymmetry, while all the
combinations of positive and negative hadrons with $z > 0.1$ are used in the
case of the dihadron asymmetry.  It has been checked, however, that the
similarity between the two different asymmetries stays the same when measuring
the asymmetries for the common hadron sample, selected with the requirement of
at least two oppositely charged hadrons produced in the primary vertex.  This
gives a strong indication that the analysing powers of the single and dihadron
channels are almost the same.

 More work has been done to understand these
similarities. Since the Collins asymmetries are the amplitudes of the sine
modulations of the Collins angles $\phi_{C^{\pm}} = \phi_{h^{\pm}} + \phi_{S} -
\pi$, where $\phi_{h^{\pm}}$ are the azimuthal angles of positive and negative
hadrons in the $\gamma^{*}$-nucleon system, the mirror symmetry suggests that in
the multi-hadrons fragmentation of the struck quark azimuthal angles of positive
and negative hadrons created in the event differ by $\approx \pi$, namely that
when a transversely polarised quark fragments, oppositely charged hadrons have
antiparallel transverse momenta.  This anti-correlation between $\phi_{h^{+}}$
and $\phi_{h^{-}}$ could be due to a local transverse momentum conservation in
the fragmentation, as it is present in the LEPTO~\cite{EdinAnders113458}
generator for spin-independent DIS.  The relevant point here is that such
correlation shows up also in the Collins fragmentation function that describes
the spin-dependent hadronisation of a transversely polarised quark $q$ into
hadrons.

\begin{figure}[!t]
\begin{center}
\includegraphics[width=0.55\textwidth]{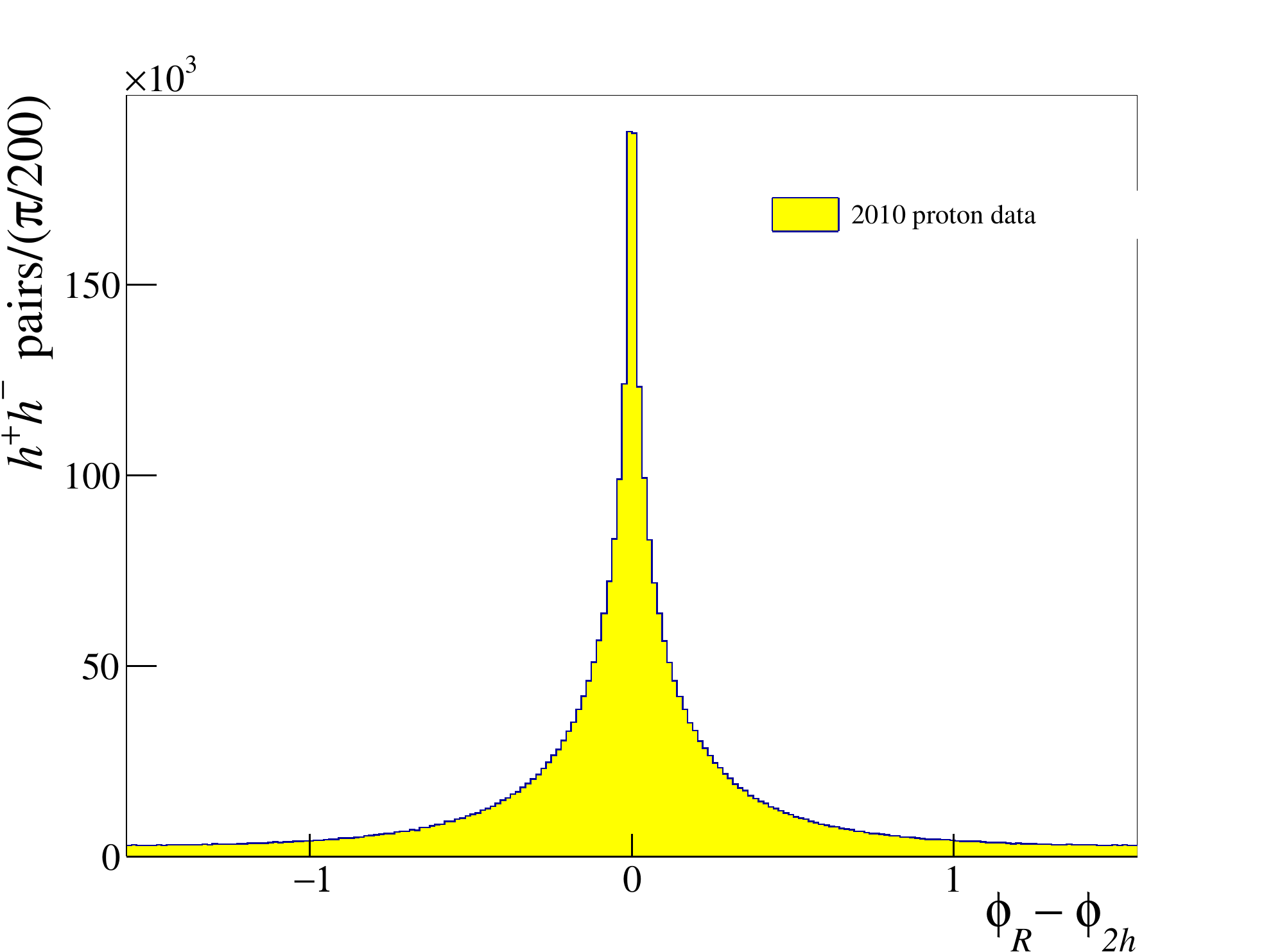}
\caption{Difference between the two dihadron angles $\phi_{R}$ and $\phi_{2h}$.}
\label{pic:phi_corr}
\end{center}
\end{figure}

\begin{figure}[!t]
\begin{center}
\includegraphics[width=0.50\textwidth]{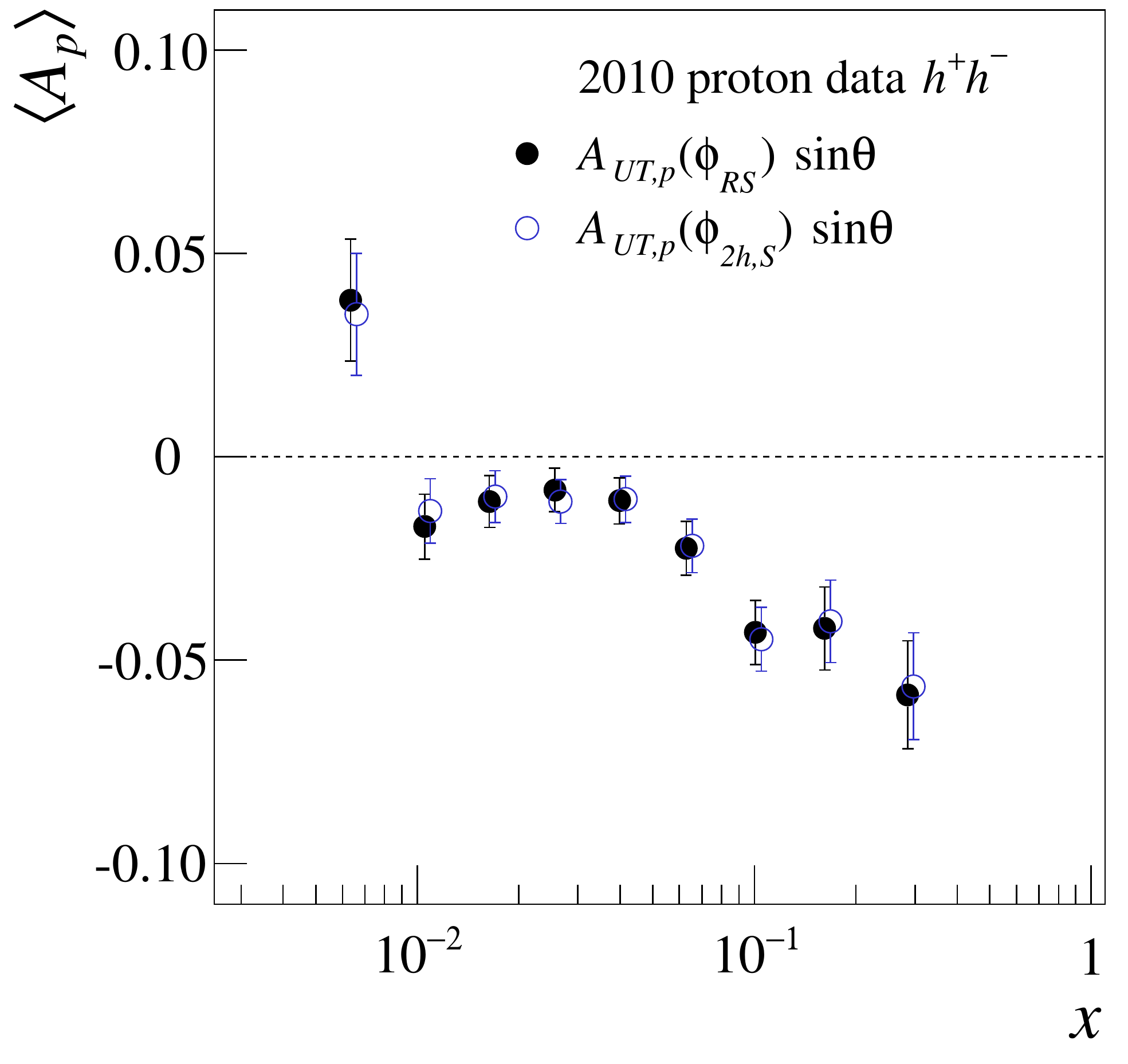}
\caption{Comparison between the dihadron asymmetry (black points) and
the Collins-like asymmetry for the dihadron (open blue points) as a function of $x$ for the $2010$ data.}
\label{pic:2h_collins_new}
\end{center}
\end{figure}

If this is the case, asymmetries correlated with the dihadrons can
also be obtained in a way different from the one described above. For each pair
of oppositely charged hadrons, using the unit vectors of their transverse
momenta, we have evaluated the angle $\textstyle \phi_{2h}$ of the vector
$\textbf{\textit{R}}_{N} = \hat{\textbf{\textit{p}}}_{T,h^{+}} -
\hat{\textbf{\textit{p}}}_{T,h^{-}}$ which is the arithmetic mean of the
azimuthal angles of the two hadrons after correcting for the discussed $\pi$
phase difference between both angles.  This azimuthal angle of the dihadron is
strongly correlated with $\phi_{R}$, as can be seen in Fig.~\ref{pic:phi_corr}
where the difference of the two angles is shown.  The same correlation is
present also in the LEPTO generator for spin-independent DIS.  Introducing the
angle $\phi_{2h,S} = \phi_{2h} - \phi_{S^{\prime}}$, one simply obtains the mean
of the Collins angle of the positive and negative hadrons (again after
correcting for the discussed $\pi$ phase difference between the two angles),
\textit{i.e.} a mean 
Collins type angle of the dihadron. The amplitudes of the modulations of $\sin
\phi_{2h,S}$, which could then be called the \textit{Collins asymmetry} for the
dihadron, are shown as a function of $x$ in Fig.~\ref{pic:2h_collins_new} for
all the $h^{+}h^{-}$ pairs with $z > 0.1$ in the $2010$ data, and compared with
the dihadron asymmetry already given in Fig.~\ref{pic:final_asyms_sys_p_2010},
where an additional cut of $p_{T} > 0.1\,\mbox{GeV}/c$ on the transverse
momentum of the individual hadrons was applied for a precise determination of
the azimuthal angles. The asymmetries are very close, hinting at a common
physical origin for the Collins mechanism and the dihadron fragmentation
function, as originally suggested in the $^{3}P_{0}$ Lund
model~\cite{Andersson198331}, in the recursive string fragmentation
model~\cite{Artru:2002pua,Artru:2010st} and in recent theoretical
work~\cite{Zhou:2011ba}.
\section {Conclusions}
\noindent
In this paper we present the results of a new measurement of the transverse spin
asymmetry in dihadron production in DIS of $160\ \mbox{GeV}/c$ muons off a
transversely polarised proton (NH$_3$) target. The measured asymmetry amplitudes
are in agreement with our previous measurement performed with data collected in
$2007$. The statistical and systematic uncertainties are considerably
reduced. The combined results show a clear signal in the $x$ range of the
valence quarks and are in agreement with a recent theoretical calculation, using
as input the transversity distribution obtained from global fits to the Collins
asymmetry. As expected, the results do not show a strong $z$ dependence. Clear
structures are exhibited as a function of the dihadrons' invariant mass, with
values compatible with zero at about $0.5\,\mbox{GeV}/c^{2}$ and a sharp fall to
$-0.05$ at the $\rho$ mass.  These new combined results will allow a more
precise extraction of the transversity distributions along the lines of the
models recently developed.  The high precision and the large kinematic range of
the COMPASS proton data allows us to compare the dihadron asymmetry and the
Collins asymmetry. In the paper we underline the striking similarity between
them and give arguments in favour of a common underlying physics mechanism, as
already suggested in the past by several authors.  In particular we show that in
our data the angle commonly used in the dihadron asymmetry analysis is very
close to the mean Collins angle of the two hadrons, and that thus the
asymmetries evaluated using the two angles turn out to be very similar.

\noindent
\section* {Acknowledgements}
This work was made possible thanks to the financial support of our funding
agencies. We also acknowledge the support of the CERN management and staff, as
well as the skills and efforts of the technicians of the collaborating
institutes.

\providecommand{\href}[2]{#2}\begingroup\raggedright
\end{document}